# Broadcast Analysis for Large Cooperative Wireless Networks

Çağatay Çapar, *Student Member, IEEE,* Dennis Goeckel, *Fellow, IEEE,* and Don Towsley *Fellow, IEEE*

*Abstract*—The capability of nodes to broadcast their message to the entire wireless network when nodes employ cooperation is considered. We employ an asymptotic analysis using an extended random network setting and show that the broadcast performance strongly depends on the path loss exponent of the medium. In particular, as the size of the random network grows, the probability of broadcast in a one-dimensional network goes to zero for path loss exponents larger than one, and goes to a nonzero value for path loss exponents less than one. In two-dimensional networks, the same behavior is observed for path loss exponents above and below two, respectively.

## I. Introduction

Cooperation among nodes is a powerful tool to improve the performance of wireless networks. A simple example of cooperation is multi-hop forwarding, where intermediate nodes transmit a source node's message along a path to a receiver which is not within the source's direct reach. More powerful forms of cooperation have emerged in recent years. For example, there has been interest in cooperative diversity, where nodes place a signal simultaneously in the same frequency band to provide spatial diversity from the source to a next-hop destination [1], [2]. Such cooperation improves link-level performance such as reducing the probability of error, outage probability, etc. [1], [3], and has been considered for improving connectivity [1] and capacity [4] in large wireless networks.

There exists a need for nodes to broadcast messages to the network. Examples include the periodic broadcast of routing updates and other control messages, and emergency signaling. It is important that a network successfully deliver these messages to the entire network, as failure to do so may block other operations and can severely impact network functionality. The important operation of broadcasting is especially challenging in a mobile ad hoc network (MANET) in which each node typically has a very limited communication range and broadcast messages have to be carried by nodes in a multi-hop fashion [5], [6]. A simple approach to broadcast in a MANET is to require each node to retransmit the message once it receives the message. However, flooding the network in this manner leads to frequent collisions and wastes network resources.

As with other network functions, and, perhaps more so, broadcast operation can be potentially enhanced by enabling a form of cooperative diversity. In particular, if the set of nodes that has decoded the message transmits using a distributed space-time code, this results in enhanced diversity against multipath fading and improved link performance rather than collisions. In addition, by combining their resources (e.g., power), cooperating nodes may be able to reach more distant nodes than would be possible without cooperation. In accordance with these ideas, several studies report improvements in broadcast performance in harsh environments [7], [8]. It is therefore important to understand the theoretical limits of broadcast performance gain that can be realized by cooperation. In this paper, we study the asymptotic limits of broadcast capabilities for large cooperative networks.

We investigate the ability of a source to transmit a message to the whole network when nodes are randomly distributed according to a Poisson point process. The source transmits the message with a given transmit power. The set of nodes which receive this signal with sufficient signal-to-noise ratio (SNR) *cooperatively* transmit the message to reach the next set of nodes which again cooperate to reach further nodes and so forth. In this manner, the broadcast message propagates through the network, and, if this wave of message transmissions reaches the entire network, the broadcast is said to be successful. In a random network, the probability of a successful broadcast is strictly less than one, as there is always a nonzero probability that the source lacks any one-hop neighbors with which to initiate broadcast in the first place. Clearly, the probability of successful broadcast monotonically decreases as the size of the network grows. In our analysis, we explore whether the broadcast probability is zero or strictly between zero and one, as the size of the network grows to infinity. The results, summarized in Table I, show that the broadcast capability of the network strongly depends on the path loss exponent. For example, in an infinite 1-D network, the broadcast probability is zero for path loss exponents larger than 1, and nonzero for path loss exponents less than 1, regardless of the node density. Note that path loss exponents $\alpha < 1$ in 1-D and $\alpha < 2$ in 2-D might seem to be of only theoretical interest, but such exponents are sometimes observed in practice [9]. The broadcast capability of cooperative networks has previously been considered in [10] under a quite different model: a finite size network where the density of nodes goes to infinity, which motivates a deterministic approach.

C. Capar and D. Goeckel are with the Department of Electrical and Computer Engineering, University of Massachusetts Amherst, Amherst, MA, 01003 USA (e-mail: ccapar@ecs.umass.edu, goeckel@ecs.umass.edu).

D. Towsley is with the Department of Computer Science, University of Massachusetts Amherst, Amherst, MA, 01003 USA (e-mail: towsley@cs.umass.edu).

TABLE I
BROADCAST PROBABILITIES OF COOPERATIVE WIRELESS NETWORKS ($\alpha$: PATH LOSS EXPONENT, $\lambda$: NODE DENSITY, $B$: EVENT OF BROADCAST, $r$: TRANSMISSION RADIUS)

| Probability of Broadcast for Extended Cooperative Networks | | | |
|---|---|---|---|
| 1-D | | 2-D | |
| $\alpha < 1$ | $0 < P(B) < 1$, $\forall \lambda > 0$ | $\alpha < 2$ | $0 < P(B) < 1$, $\forall \lambda > 0$ |
| $\alpha = 1$ | $0 < P(B) < 1$, $\forall \lambda > 1, r = 1$ | $\alpha = 2$ | $0 < P(B) < 1$, $\forall \lambda > 4/\pi, r = 1$ |
| $\alpha > 1$ | $P(B) = 0$ | $\alpha > 2$ | $P(B) = 0$ |

The rest of the paper is organized as follows: Section II describes the network assumptions and introduces the cooperative communication model. In Section III, we establish results on broadcast performance of cooperative wireless networks summarized in Table I. Section IV is the conclusion.

## II. COOPERATIVE NETWORK MODEL

We assume an extended wireless network, where nodes are randomly distributed in an infinite region according to a Poisson point process with node density $\lambda$. Each node is assumed to transmit with peak power $P_t$, which allows it to communicate, without cooperation, to nodes within transmission radius $r$. A node's transmission radius is defined to be the range within which other nodes can receive its signal with a power above a specified decoding threshold, $\tau$, which allows a receiver to satisfy a minimum signal-to-noise ratio (SNR) for physical layer functionality so that the two nodes are *connected*. With these definitions, $r$ is given by:

$$P_t r^{-\alpha} = \tau, \qquad (1)$$

where $\alpha$ is the path loss coefficient which determines the rate at which the received power decays with distance.

When two or more nodes cooperate, they simultaneously transmit the same message such that they can reach a greater distance than they would otherwise reach without cooperation. In this work, we assume cooperation provides power summing at the receiver. More formally, the condition for a cooperating set of nodes $\Omega$ to reach a node $k$ is

$$P_t \sum_{j \in \Omega} (d_{j,k})^{-\alpha} \geqslant \tau, \qquad (2)$$

where $d_{j,k}$ is the distance between nodes $j$ and $k$.

We assume that the source initiates cooperative broadcast by transmitting the message, which is heard by nodes within the source's transmit range. In the second step, those nodes that have just received the message transmit cooperatively and reach a further set of nodes. In successive steps, the set of nodes that have received the message from the previous step cooperatively transmit. As in the maximum multihop diversity case in [10], we assume that a receiving node accumulates power from all previous steps. Hence, a node $k$ can successfully decode the broadcast message, if this accumulated power satisfies (2), with $\Omega$ being the set of nodes that have previously decoded the message. If a message transmitted from a given node reaches all of the nodes in the network, then broadcast is said to be achieved.

## III. BROADCAST ANALYSIS

With our assumptions on the network given in Section II, there is always a nonzero probability that there are no nodes in the transmission range of the source, which means the broadcast cannot even be initiated. Hence, our analysis focuses on determining whether the broadcast probability is strictly larger than zero or not.

The problem considered is closely related to *percolation*, which has been extensively studied for (non-cooperative) wireless networks [11]. For an infinite network, percolation is said to occur if there exists a connected cluster with an infinite number of nodes. More formally, for a random graph, if the probability that a vertex (e.g., a node at the origin) connects to an infinite number of vertices is strictly larger than zero, this graph is said to percolate. In order to study the properties such as percolation, full connectivity, capacity, etc., an asymptotic analysis of wireless networks is usually done in one of two ways [12]: one can fix the size of the network and let the number of nodes go to infinity (dense network), or one can let the size of the network grow for some fixed node density (extended network). These two models can be seen as two different views at the same network scaling as long as some physical modeling issues are addressed (e.g., power law path loss assumption causing a singularity for dense networks [13]). Here, we study the broadcast capability of cooperative wireless networks in an extended network setting.

In the following analysis, without loss of generality, we assume that $\tau = P_t$, which makes the transmission radius $r = 1$ in (1). Some of the results (Theorems 1 (first part), 2, 3 (first part)) are for all node densities; therefore, those results are valid for any $r > 0$. The results in Theorems 1 (second part) and 3 (second part) are valid for $\lambda r > 1$ and $\lambda r > 4/\pi$.

### A. 1-D Networks

*Theorem 1:* In a 1-D extended cooperative wireless network, a node can broadcast its message to the entire network with nonzero probability (i) for any node density $\lambda > 0$ and any path loss exponent $\alpha < 1$, (ii) for any node density $\lambda > 1$ and path loss exponent $\alpha = 1$.

*Proof:* Assume there is a node at the origin, and consider broadcast in the positive direction on the line. Showing a nonzero probability of broadcast in the positive direction implies a nonzero broadcast probability for the entire network. Divide the line into intervals $\{L_k\}$ as given in Fig. 1. Notice that the length of interval $L_k$ is equal to $k$. In the following, we describe an event corresponding to a sufficient number of nodes in each interval for broadcast, and then show that this event occurs with positive probability.

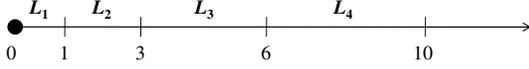

Fig. 1. Division of the line used in the proof of Theorem 1.

Define the following events:

$B_1$: the event that interval $L_1 = (0,1]$ contains at least $\lceil 2^\alpha \rceil$ nodes. $B_2$: the event that interval $L_2 = (1,3]$ contains at least $\lceil 3^\alpha \rceil$ nodes, $\cdots$, $B_k$: the event that interval $L_k$ contains at least $\lceil (k+1)^\alpha \rceil$ nodes. Let the random variable $n_k$ be the number of nodes in interval $L_k$. With this definition,

$$P(B_k) = P(n_k \geq \lceil (k+1)^\alpha \rceil), \quad \forall k \in \{1, 2, \cdots\}.$$

The node at the origin can reach nodes within $L_1$ without cooperation. $B_1$ guarantees that the message can be broadcast to nodes within $L_1 \cup L_2$. Similarly if $\bigcap_{k=1}^n B_k$ occurs, this guarantees broadcast in $\bigcup_{k=1}^{n+1} L_k$. This can be seen by lower bounding the power received at the rightmost end of $\bigcup_{k=1}^{n+1} L_k$ by assuming all nodes in $\bigcup_{k=1}^n L_k$ are located at the origin and recalling $\alpha \leqslant 1$. With all nodes assumed to be located at the origin, the received power would be:

$$\frac{P_t(1 + \sum_{k=1}^n \lceil (k+1)^\alpha \rceil)}{|\bigcup_{k=1}^{n+1} L_k|^\alpha} \geq \frac{P_t(\sum_{k=1}^{n+1} k^\alpha)}{(\sum_{k=1}^{n+1} k)^\alpha} \geq \tau. \quad (3)$$

Note the simplifying assumption that $P_t = \tau$. The next step is to show that the event $\bigcap_{k=1}^\infty B_k$, which enables broadcast, has nonzero probability.

Consider $\lambda > 0$ for $\alpha < 1$, and $\lambda > 1$ for $\alpha = 1$. The number of nodes $n_k$ in the interval $L_k$ has expected value $k\lambda$. Let $N$ be the smallest integer such that this expected value is greater than the number of nodes required for $B_N$ to occur; that is $N\lambda > \lceil (N+1)^\alpha \rceil$. It is always possible to find such $N$ because as $k \to \infty$, $\lceil (k+1)^\alpha \rceil / k \to 0$ for $\alpha < 1$, and $\lceil (k+1)^\alpha \rceil / k \to 1$ for $\alpha = 1$. Now write

$$P(B^*) \geq P(\bigcap_{k=1}^\infty B_k) = \prod_{i=1}^{N-1} P(B_i) \prod_{k=N}^\infty P(B_k), \quad (4)$$

where $B^*$ denotes the event that broadcast in the positive direction on the line happens. Clearly $\prod_{i=1}^{N-1} P(B_i) > 0$. Next consider $\prod_N^\infty P(B_k)$. Let $\delta \in (0,1)$ be defined as:

$$(1-\delta)E(n_N) = (1-\delta)\lambda N = \lceil (N+1)^\alpha \rceil \quad (5)$$

Using a Chernoff bound, we find a lower bound for $P(B_N)$.

$$\begin{aligned} P(B_N) &= P(n_N \geq \lceil (N+1)^\alpha \rceil) \\ &= P(n_N \geq (1-\delta)\lambda N) \\ &\geq 1 - \exp(-\lambda N \delta^2 / 2). \end{aligned} \quad (6)$$

The final step is to find lower bounds for $P(B_k)$ for $k > N$. As $k$ increases, the ratio of the required number of nodes in interval $L_k$ to the expected number $k\lambda$ gets smaller. Thus, using the same $\delta$, which is constant given $N$ as above,

$$\begin{aligned} P(B_k) &= P(n_k \geq \lceil (k+1)^\alpha \rceil) \\ &\geq 1 - \exp(-\lambda k \delta^2 / 2) \quad \text{for } k \geq N. \end{aligned} \quad (7)$$

Then,

$$\prod_{k=N}^\infty P(B_k) \geq \prod_{k=N}^\infty (1 - \exp(-\lambda k \delta^2 / 2)) > 0. \quad (8)$$

The last inequality above can be seen by noting that $\sum_{k=N}^\infty \exp(-\lambda k \delta^2 / 2)$ is a convergent series and hence, the infinite product is convergent [14]. Finally,

$$P(B^*) \geq \prod_{i=1}^{N-1} P(B_i) \prod_{k=N}^\infty P(B_k) > 0, \quad (9)$$

and $P(B) > 0$, where $B$ is the event of broadcast to the entire line. ∎

Note that Theorem 1 states that a randomly selected node can broadcast to the entire network with nonzero probability. This implies that, with probability one, there exists a node on the line which can broadcast to the entire network. This argument can be shown by considering the network as a disjoint union of finite size regions. For arbitrarily high probability, these regions can be selected large enough, so that if a node can broadcast to an entire region, it can broadcast to the rest of the network with that probability. The result follows by noticing that, with arbitrarily high probability, there exists a node inside a finite union of these regions which can broadcast to its entire region.

*Theorem 2:* In a 1-D extended cooperative wireless network, the probability that a node can broadcast its message to the entire network is zero for any node density $\lambda > 0$ for path loss exponent $\alpha > 1$.

*Proof:* Without loss of generality, assume that $P_t = \tau = 1$. For some $t \in \mathbb{R}$, consider nodes distributed in $(-\infty, t)$ according to a Poisson point process. Let $x$ represent a realization of this random process, and let $\mathcal{X}$ be the set of all realizations. For some $l > 0$, a node to the left of $t$ can send a broadcast message to point $(t+l)$ on the line only if, at some time in the execution of the broadcast, it belongs to a connected cluster $\Omega$ such that

$$\sum_{k \in \Omega} \frac{1}{(t+l-x_k)^\alpha} \geq 1, \quad (10)$$

where $x_k$ is the location of node $k$ in $\Omega$. Let $A \subset \mathcal{X}$ denote the set of all such realizations.

Next, let $B \subset \mathcal{X}$ be the set of realizations such that if all the nodes in $(-\infty, t)$ transmit simultaneously, the received power at $(t+l)$ is larger than the threshold; i.e.,

$$B = \{x \in \mathcal{X} | \sum_k \frac{1}{(t+l-x_k)^\alpha} \geq 1\} \quad (11)$$

where, $x_k$ is the location of node $k$ for the realization $x \in \mathcal{X}$. For any $x \in A$, clearly $x \in B$, and hence $A \subset B$. So, $I_A(x) \leq I_B(x), \forall x \in \mathcal{X}$, where $I_A(\cdot), I_B(\cdot)$ denote the indicator functions of the sets $A, B$, respectively. The probability that a node to the left of $t$ can broadcast to a node at $(t+l)$ is $E[I_A]$, which, by the monotonicity of integration, is upper bounded by $E[I_B]$. Therefore, for any $l > 0$ and $t \in \mathbb{R}$, the broadcast probability at $(t + l)$ by nodes in $(-\infty, t)$ is upper bounded by the probability that combined power from $(-\infty, t)$ can reach $(t + l)$.

Now, consider a node $j$ at $x_j$, and assume nodes are distributed according to a Poisson point process to the left of $x_j$ in $(-\infty, x_j)$. For some $l > 0$, consider the power received at $j$ when all the nodes to the left of $(x_j - l)$ transmit simultaneously:

$$Y = \sum_{x_k \in (-\infty, x_j - l)} \frac{1}{(d_{k,j})^\alpha},$$

where $d_{k,j}$ is the distance between nodes $j$ and $k$.

When $\alpha > 1$, the random variable $Y$ has a finite mean, $\mu = E(Y) < \infty$. Next, define $Y(d)$ for an *integer* $d > l$ as

$$Y(d) = \sum_{x_k \in [x_j - d, x_j - l)} \frac{1}{(d_{k,j})^\alpha}. \quad (12)$$

For any sample point for which $Y$ converges, $Y(d)$ is non-negative and non-decreasing with $d$, and, hence

$$E(Y(d)) \to E(Y), \quad d \to \infty \quad (13)$$

by the monotone convergence theorem. Then for arbitrarily small $\varepsilon/2 > 0$, there exists a $d^*$ such that for $d \geq d^*$, $E(Y) - E(Y(d)) < \varepsilon/2$. Define:

$$Z(d) = Y - Y(d) = \sum_{x_k \in (-\infty, x_j - d)} \frac{1}{(d_{k,j})^\alpha}.$$

By the Markov inequality,

$$P(Z(d^*) > 1) \leq E(Z(d^*)) < \frac{\varepsilon}{2}. \quad (14)$$

Hence, with probability larger than $1 - \varepsilon/2$, a node with no neighbors within $d^*$ to its left cannot be reached by combined power from all the nodes to its left. Then, as shown above, the probability that such a node can receive a broadcast message by cooperation of nodes to its left is less than $\varepsilon/2$.

Next, consider a 1-D network where nodes are distributed in $(-\infty, \infty)$ according to a Poisson point process with density $\lambda > 0$. Starting from the origin and moving to the left, consider the first gap which is larger than $d^*$. For any $\varepsilon > 0$, such a gap exists within the first $N$ gaps with probability

$$1 - (1 - \exp(-\lambda d^*))^N > 1 - \varepsilon/2$$

for some integer $N$. Consider the node at the right end of this gap. By (14), the probability that a broadcast message can be delivered to this node by nodes to its left is less than $\varepsilon/2$. Hence, probability of broadcast to the whole network can be upper bounded by an arbitrarily small number $\varepsilon > 0$. ∎

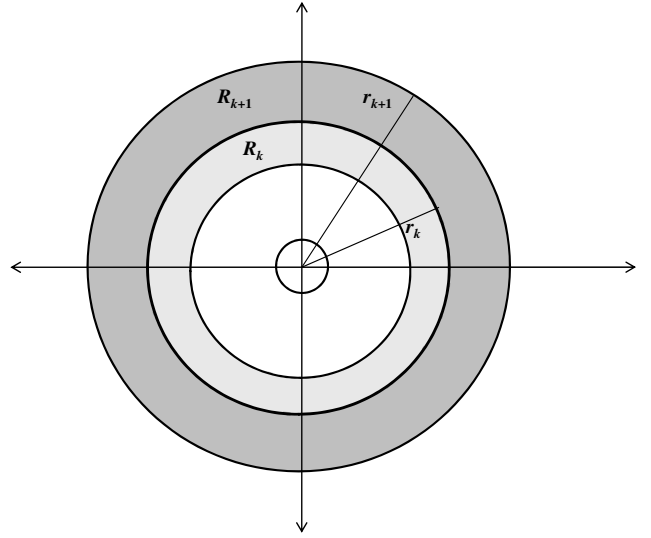

Fig. 2. Division of $\mathbb{R}^2$ into rings as used in proof of Theorem 3.

### B. 2-D Networks

*Theorem 3:* In a 2-D extended cooperative wireless network, a node can broadcast its message to the entire network with nonzero probability (i) for any node density $\lambda > 0$ and any path loss exponent $\alpha < 2$, (ii) for any node density $\lambda > \frac{4}{\pi}$ and path loss exponent $\alpha = 2$.

*Proof:* Consider a division of $\mathbb{R}^2$ into rings as given in Fig. 2. We are interested in the probability that a node at the origin can broadcast its message to the entire network. Similar to the 1-D case, we describe an event corresponding to a sufficient number of nodes in each ring for broadcast, and then show that this event occurs with positive probability.

Let $R_k$ denote the $k$th ring and $r_k = \sqrt{1 + 2 + \cdots + k}$ be the radius of the circular area including the first $k$ rings. Notice that the $k$th ring has area $\pi k$. Let the random variable $n_k$ denote the number of nodes in the $k$th ring. We define the following events: $B_1$ is the event that $R_1$ contains at least $\lceil 2^\alpha (1 + 2^{\alpha/2}) \rceil$ nodes, and for $k \geq 2$, $B_k$ is the event that $R_k$ contains at least $\lceil 2^\alpha (k+1)^{\alpha/2} \rceil$ nodes.

The node at the origin can reach any node within the first ring (which is in fact a disk of radius 1) without cooperation. The event $\bigcap_{k=1}^{n} B_k$ guarantees broadcast in $\bigcup_{k=1}^{n+1} R_k$. This can be seen by lower bounding the power coming from the first $k$ rings received at a point located at distance $r_{k+1}$ from the origin. If all nodes in the first $k$ rings are assumed to be located at the maximum possible distance from this point (i.e., at distance $r_k + r_{k+1}$ each), the power received would be

$$P_t \frac{(1 + n_1 + n_2 + \cdots + n_k)}{(r_k + r_{k+1})^\alpha} \tag{15}$$

$$\geq P_t \frac{2^\alpha (1 + 2^{\alpha/2} + \cdots + (k+1)^{\alpha/2})}{(2 r_{k+1})^\alpha} \tag{16}$$

$$= P_t \frac{2^\alpha \sum_{i=1}^{k+1} i^{\alpha/2}}{2^\alpha (\sum_{i=1}^{k+1} i)^{\alpha/2}} \geq \tau, \quad \text{for } \alpha \leq 2. \tag{17}$$

Therefore, the transmission power from the first $k$ rings suffices to reach the $(k+1)$th ring.

We next show that $P(\cap_{k=1}^\infty B_k) > 0$. Note that the number of nodes in the $k$th ring has expected value $E[n_k] = \pi \lambda k$. At each step for $k \geq 2$, we require $\lceil 2^\alpha (k+1)^{\alpha/2} \rceil$ nodes in $R_k$. Let $N$ be the smallest integer such that $\pi \lambda N > \lceil 2^\alpha (N+1)^{\alpha/2} \rceil$. It is always possible to find such an $N$ because as $k \to \infty$, $\lceil 2^\alpha (k+1)^{\alpha/2} \rceil / \pi k \to 0$ for $\alpha < 2$, and $\lceil 2^\alpha (k+1)^{\alpha/2} \rceil / \pi k \to 4/\pi$ for $\alpha = 2$. Let $B$ be the event that broadcast happens. Then

$$P(B) \geq P(\bigcap_{k=1}^\infty B_k) = \prod_{i=1}^{N-1} P(B_i) \prod_{k=N}^\infty P(B_k). \tag{18}$$

Clearly $\prod_{i=1}^{N-1} P(B_i) > 0$. Next consider $\prod_{k=N}^\infty P(B_k)$. Let $\delta \in (0, 1)$ be defined as:

$$(1 - \delta) E(n_N) = (1 - \delta) \pi \lambda N = \lceil 2^\alpha (N+1)^{\alpha/2} \rceil \tag{19}$$

Using a Chernoff bound, we find a lower bound for $P(B_N)$.

$$\begin{aligned} P(B_N) &= P(n_N \geq \lceil (N+1)^{\alpha/2} \rceil) \\ &= P(n_N \geq (1 - \delta) \pi \lambda N) \\ &\geq 1 - \exp(-\pi \lambda N \delta^2 / 2) \end{aligned} \tag{20}$$

The final step is to lower bound $P(B_k)$ for $k > N$. As $k$ increases, the ratio of the required number of nodes in the ring $R_k$ to the expected number $\pi \lambda k$ gets smaller. Thus, using the same $\delta$, which is constant given $N$ as above,

$$\begin{aligned} P(B_k) &= P(n_k \geq \lceil 2^\alpha (k+1)^{\alpha/2} \rceil) \\ &\geq 1 - \exp(-\pi \lambda k \delta^2 / 2) \quad \text{for } k \geq N. \end{aligned} \tag{21}$$

Then,

$$\prod_{k=N}^\infty P(B_k) \geq \prod_{k=N}^\infty (1 - \exp(-\pi \lambda k \delta^2 / 2)) > 0, \tag{22}$$

which implies

$$P(B) \geq \prod_{i=1}^{N-1} P(B_i) \prod_{k=N}^\infty P(B_k) > 0.$$

∎

*Theorem 4:* In a 2-D extended cooperative wireless network, the probability that a node can broadcast its message to the entire network is zero for any node density $\lambda > 0$ for path loss exponent $\alpha > 2$.

*Proof:* The proof is very similar to the proof of Theorem 2. It is based on the idea that there exists a critical distance $r^* < \infty$ such that, with high probability, the network cannot deliver a broadcast message to a node that has no neighbors within a radius of $r^*$. As the size of the network grows, it includes such an isolated node with high probability. ∎

Thus, in a 2-D extended network with $\alpha > 2$, broadcast to the entire network is not possible. However, in contrast to a 1-D network with $\alpha > 1$, it is possible (with nonzero probability) for a node to broadcast its message to an *infinite* number of nodes in a 2-D extended network with $\alpha > 2$. This result follows from Theorem 4.3 of [15].

*C. Comparison with Continuum Analysis*

A theoretical analysis with the same goal of considering broadcast performance in cooperative wireless networks has previously been considered in [10] for a dense network. The key element of the analysis in [10] (and similarly in [16], [17], [18]) is that the random network is approximated by a deterministic *continuum model*, where it is assumed that the transmit power is distributed to the entire network as a continuum as opposed to separate randomly placed nodes in discrete locations with some nonzero transmit power. For a 2-D dense network where the transmit power coming from the cooperating nodes is assumed to be summed at the receiver, [10] shows that broadcast performance depends on what scale *multihop diversity* is exploited. In multihop diversity, power is accumulated from $m \geq 1$ previous levels. If, as assumed here, power is accumulated from all previous levels ($m = \infty$), it is shown (for path loss exponent $\alpha = 2$) that the message is broadcast to the whole network [10]. This result can be easily extended to arbitrary path loss exponents in a 1-D network by observing that in the continuum limit, if nodes in a region of size 1 can reach nodes within a region of size $(1 + \varepsilon_1), \varepsilon_1 > 0$, in the next level, broadcast will reach a region of size $(1 + \varepsilon_1 + \varepsilon_2)$ with $\varepsilon_2 > \varepsilon_1$. Hence, the broadcast region grows to infinity for all path loss exponents and power densities. The same result follows analogously in 2-D.

Thus, the results of [10] appear to be quite at odds with the results derived here. However, the dichotomy can be explained by carefully considering the assumptions of the two papers. In the continuum analysis of [10], one models the random network in the limit of high node densities, where the distribution of nodes becomes deterministic, and then checks whether broadcast is possible or not. As noted above, under the $m = \infty$ assumption, this results in broadcast with probability one to the entire network, even in the limit of very large networks, because of the deterministic uniformity of the node distribution. Here, in contrast, we consider fixed node densities, with the associated randomness of node distribution, in the limit of large network size. Because of the randomness in the node locations, for $\alpha > 1$ (1-D) or $\alpha > 2$ (2-D) it is very likely that one will find an isolated node if the network is large enough, as formally described in the proofs of Theorems 2 and 4. Hence, the probability of broadcast is zero regardless of the node density $\lambda$. Hence, it is clear that one must be careful in choosing the appropriate model for a given application. The model of [10] has been successfully employed in numerous works [16], [19], [20], but our results would suggest caution in its application to very large random networks which require many hops for broadcast to reach the entire network.

## IV. Conclusion

We analyze the theoretical limits of node broadcast in a cooperative wireless network. For a network where nodes are distributed randomly to a region, we calculate the probability that a node can broadcast its message to the whole network when the size of the network grows to infinity. Using an exact discrete model, we show that broadcast performance of the cooperative network strongly depends on the path loss exponent, and that there is zero probability of broadcast for a large range of path loss exponents regardless of the node density.

## Acknowledgment

This work was supported in part by the National Science Foundation under Grants CNS-0721861 and CNS-1018464, and by the U.S. Army Research Laboratory and the U.K. Ministry of Defence and was accomplished under Agreement Number W911NF-06-3-0001. The views and conclusions contained in this document are those of the author(s) and should not be interpreted as representing the official policies, either expressed or implied, of the U.S. Army Research Laboratory, the U.S. Government, the U.K. Ministry of Defence or the U.K. Government.